\documentclass[12pt,
 reprint,
 amsmath,
 amssymb,
 aps,
]{revtex4}

\usepackage{braket}
\usepackage{graphicx}
\usepackage{dcolumn}
\usepackage{bm}
\usepackage{tikz}
\usepackage{amsmath}
\usepackage[utf8]{inputenc}

\usepackage{chemformula}

\usepackage{setspace} 

\begin{document}

\UseRawInputEncoding

\preprint{APS/123-QED}

\title{Role of Kinetic Exchange and Coulomb Interaction in Bonding of Hydrogen Molecular Systems and Excited States}

\author{Maciej Hendzel}
\email{maciej.hendzel@doctoral.uj.edu.pl}
\author{J\'ozef Spa\l{}ek}
\email{jozef.spalek@uj.edu.pl}
 \affiliation{Institute of Theoretical Physics, Jagiellonian University,\\ ul.~\L{}ojasiewicza 11, PL-30-348 Krak\'{o}w, Poland}

\date{\today}

\begin{abstract}
We present a detailed investigation of the electronic structure and bonding 
characteristics of hydrogen-based molecular systems (\ch{H2+}, \ch{H2}, \ch{H2-}) 
using the Exact Diagonalization Ab Initio (EDABI) approach within the framework of 
combined first- and second-quantization. By analyzing the relative contributions of 
kinetic exchange and effective Coulomb interactions, we provide a comprehensive 
understanding of covalency, 
atomicity, and ionicity as a function of interatomic distances. Our approach leverages 
exact solutions of the extended Heitler-London model to quantify these 
interactions, extending the analysis to the discussion of properties of excited states 
and the dissociation limit to 
these molecules. The findings reveal significant differences in bonding 
characteristics, particularly highlighting the stability and bonding nature of the 
neutral \ch{H2} molecule compared to its ionic counterparts. This study not only 
enhances an understanding of molecular interactions in hydrogen systems but also 
demonstrates the potential of the EDABI approach in developing more accurate 
computational models in quantum chemistry. 
\end{abstract}

\maketitle

\clearpage
\newpage
Fundamental understanding of molecular electronic structure and bonding is crucial for 
various applications in chemistry and physics. Despite numerous remarkable works 
published, the discussion on selected aspects of chemical bonding remains intense, 
even for relatively simple systems like diatomic molecules \cite{pendas2022role}. In 
our view, this ongoing debate is related to the advancement of numerical methods, 
which remain interpretative challenges despite achieving highly accurate 
numerical results with the appropriate level 
of complexity and sophistication.

To provide a brief outline of this discussion, it is worth mentioning works related to 
the ambiguity of the concepts of ionicity and covalency in chemical bonds 
\cite{Pendas, Fugel}, as well as attempts to explain the covalency of the bond in the 
hydrogen molecule and other diatomic molecules \cite{Bacskay2017, ChenSeniority, 
Levine, pendas2022role}. Our contribution to this discussion has been to provide the 
complementary characteristics of the chemical bond such as the atomicity and true 
covalency \cite{Hendzel2022, hendzel2022toward} or our newest paper about entanglement 
correlations as the alternative bonding characteristic  \cite{JS_Hen_arxiv}.

In this paper, we use the EDABI approach to investigate the detailed electronic 
structure of hydrogen-based molecular systems, specifically \ch{H2+}, \ch{H2}, and 
\ch{H2-}. We employ the formalism of second quantization to express the Hamiltonian 
and wavefunctions, enabling a clear and concise representation of the electronic 
states. The Hamiltonian we consider includes all two-electron interactions in the Fock space that appear in the two-orbital model of the Heitler-London type. 

One of the key aspects we explore is the concept of the bonding characteristics, 
namely covalency, atomicity, and ionicity, within the context of molecular 
dissociation. By examining the bonding factors as a function of interatomic distance, 
we provide an insight into the nature of chemical bonding and the evolution between 
different physical regimes. The analysis is based on the exact solutions of the 
extended Heitler-London model, allowing us to rigorously quantify the contributions of 
kinetic exchange interactions and effective Coulomb interactions to the overall 
bonding. Furthermore, we extend our investigation to the excited states of hydrogen 
molecules and discuss how they compare to the ground state.

\section{Method}

The total electronic energy of a system is expressed 
as the sum of the kinetic interaction part energy. The kinetic energy term in Hartree-Fock corresponds to the 
average kinetic energy of non-interacting electrons in the effective potential 
generated by the other electrons. On the other hand, the exchange kinetic energy 
refers to the contribution to the total kinetic energy that arises from the 
exchange interaction between identical particles, such as electrons. This is a 
quantum mechanical effect resulting from the indistinguishability of identical 
particles.

On the other hand, methods such as Configuration Interaction (CI), Coupled Cluster (CC), 
and Exact Diagonalization Ab Initio (EDABI)

To single out those exchange and other contributions we start from two general form of the Hamiltonian in the Fock space

\begin{align}
    \hat{\mathcal{H}}= \sum_{ij\sigma} t_{ij} \hat{a}_{i\sigma}^{\dag}\,\hat{a}_{j\sigma} +
    \frac{1}{2} \sum_{\substack{ijkl \\ \sigma\sigma'}} 
    V_{\substack{ijkl \\ \sigma\sigma'}}\, \hat{a}_{i\sigma}^{\dag}\,\hat{a}_{j\sigma'}^{\dag}\, 
    \hat{a}_{l\sigma'}\,\hat{a}_{k\sigma},
    \label{Hamiltonian_fetter-walecka}
\end{align}
where 
\begin{align}
    t_{ij\sigma}\equiv \braket{ \phi_{i\sigma} |\mathcal{H}_{1}| \phi_{j\sigma} }, 
\end{align}
and
\begin{align}
    V_{ijkl}= \braket{ \phi_{i\sigma}\phi_{j\sigma'} |V|  \phi_{k\sigma'}\phi_{l\sigma}}.
    \label{t_and_V}
\end{align}

\noindent
In these expressions, $t_{ij\sigma}$ represents the one-electron integrals, 
including kinetic energy and nuclear attraction, while $V_{ijkl}$ are the two-
electron integrals representing electron-electron repulsion. The operators 
$\hat{a}_{i\sigma}^{\dag}$ and $\hat{a}_{j\sigma}$ are the creation and 
annihilation operators for electrons in orbitals $i$ and $j$ with spin $\sigma$.

The general $N$-particle state $\ket{\Phi_{N}}$ in Fock space can be related to the corresponding $N$-particle wavefunction $\Psi_{\alpha}(\textbf{r}_1, \ldots, \textbf{r}_N)$ in Hilbert space as follows \cite{Robertson}
\begin{align}
\ket{\Phi_N} = \frac{1}{\sqrt{N!}}
 \int d^3\textbf{r}_1 \ldots d^3\textbf{r}_N \,
 \Psi_{N} (\textbf{r}_1, \ldots, \textbf{r}_N) \notag \\
 \hat{\Psi}_{\sigma_1}^{\dag}(\textbf{r}_1) \ldots \hat{\Psi}_{\sigma_N}^{\dag}(\textbf{r}_N) \ket{0}.
\end{align}

\noindent
where the field operator $\hat{\Psi}(\textbf{r})$ is here approximated using a finite number $M$ of wavefunctions $\{w_i(\textbf{r})\}$

\begin{align}
\hat{\Psi}(\textbf{r}) &\simeq \sum_{i=1}^{M} w_i(\textbf{r}) \hat{a}_{i},
\end{align}

\noindent
where the Wannier-Mulliken function is taken in the following form of molecular orbitals

\begin{align}
    w_i(\textbf{r}) = \beta (\psi_i(\textbf{r}) -\gamma\psi_{i+1}(\textbf{r})).
\end{align}

\noindent
Such functions are normalized and orthogonal, which leads to the following expressions for $\beta$ and $\gamma$ 

\begin{align}
\beta &= \frac{1}{\sqrt{2}} \sqrt{\frac{1 + \sqrt{1 - S^2}}{1 - S^2}} \\
\gamma &= \frac{S}{1 + \sqrt{1 - S^2}}, 
\end{align}

\noindent
where $S = \braket{\psi_1(\textbf{\textbf{r}})|\psi_2(\textbf{r})}$ denotes the overlap integral.

The approximate $N$-particle wavefunction is then

\begin{align}
&\Psi_{\alpha}(\textbf{r}_1, \ldots, \textbf{r}_N) = \notag \\
&\frac{1}{\sqrt{N!}} \sum_{i_1, \ldots, i_N = 1}^{M} \langle 0 | \hat{a}_{i_N} \ldots \hat{a}_{i_1} | \Phi_{N} \rangle w_{i_1}(\textbf{r}_1) \ldots w_{i_N}(\textbf{r}_N). \label{calc_wav}
\end{align}

In Fock space, the $N$-particle state is expressed as

\begin{align}
|\Phi_{N}\rangle &= \frac{1}{\sqrt{N!}} \sum_{j_1, \ldots, j_N = 1}^{M} C_{j_1 \ldots j_N} \hat{a}_{j_1}^{\dag} \ldots \hat{a}_{j_N}^{\dag} | 0 \rangle.
\end{align}

Substituting this into the $N$-particle wavefunction gives

\begin{align}
&\Psi_{\alpha}(\textbf{r}_1, \ldots, \textbf{r}_N) = \notag \\
&\frac{1}{\sqrt{N!}} \sum_{i_1, \ldots, i_N = 1}^{M} C_{i_1 \ldots i_N}(A,S) 
w_{i_1}(\textbf{r}_1) \ldots w_{i_N}(\textbf{r}_N).
\end{align}

where $C_{i_1 \ldots i_N}(A,S)$ are the coefficients determined by the diagonalization. This represents the Configurational Interaction wavefunction for $N$ particles distributed among $M$ states, with $A$ and $S$ indicating antisymmetrization (Slater determinant) for fermions or symmetrization for bosons. In CI, the wavefunction is expressed as a linear combination of Slater determinants, and the coefficients $C_{i_1 \ldots i_N}(A,S)$ are optimized variationally. 

EDABI also involves diagonalizing the Hamiltonian matrix to find the exact wave function within a given basis set. The CI approach shares this feature but focuses on a finite basis set of $M$ states, neglecting highly excited states to simplify the problem.

The CI method used here complements these approaches by determining the coefficients 
$C_{i_1 \ldots i_N}$ through diagonalization in Fock space, potentially offering a 
more computationally efficient alternative while still capturing essential 
correlations in the system.

\begin{figure}[t]
    \centering
    \includegraphics[width=0.4\textwidth]{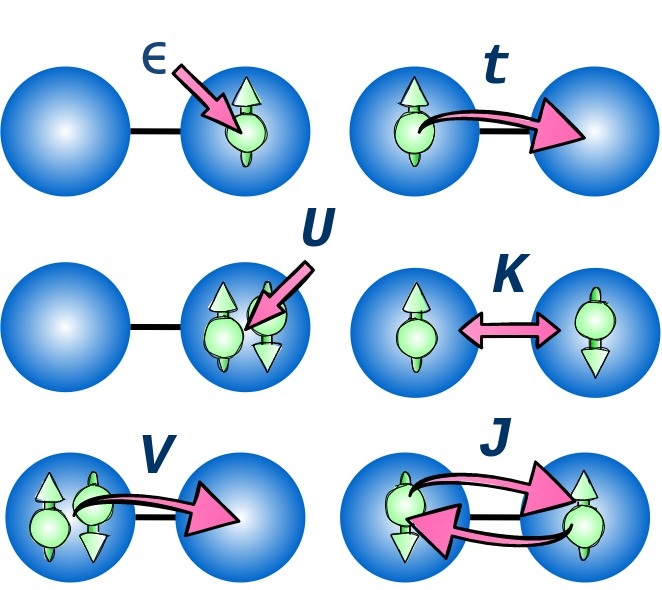}
    \caption{Schematic illustration of various interaction processes within molecular systems, highlighting several key elements: single-particle excitation, electron hopping between adjacent sites, on-site Coulomb repulsion between electrons, direct exchange interaction between sites, inter-site Coulomb interaction, and indirect exchange interaction (superexchange). The green arrows represent particles with spin, while the pink arrows depict the pathways of these interactions.}
    \label{fig:integrals}
\end{figure}

The full electronic Hamiltonian for the two-orbital model was derived and discussed 
earlier \cite{SpalJul, Hendzel2022, hendzel2022toward}. Here we present schematically 
all physical processes which are involved in this Hamiltonian in Fig.~\ref{fig:integrals}.
In the figure $\epsilon_a$ denotes a single-particle energy of 
the electron in an atomic state and it is shown as an isolated electron with spin.  The second microscopic parameter: hopping amplitude $t$ represents electron hopping between two neighboring atoms. It is represented schematically in the 
Fig.~\ref{fig:integrals} as an electron (green arrow) moving from one site to 
another. The $U$ term accounts for the Coulomb repulsion between electrons on the 
same site. It is indicated in the Figure by a purple arrow representing the 
repulsion between two electrons at the same site (green arrows indicating 
opposing spins). The $K$ term represents the Coulomb interaction between 
electrons on different sites. The figure illustrates this as a repulsive force 
between electrons on adjacent sites (straight purple double arrow between two 
green arrows). 
The exchange integral $J$ is shown in the figure a green arrows forming loops between 
two atoms occupied by two electrons with opposite spins suggesting the interaction involving an exchange of spin orientation. $V$ is the so-called 
correlated hopping marked as a purple arrow from one atom to its neighbor indicating 
the hopping of an electron pair. 

\section{Results}

We discuss here results for hydrogen-based molecular systems like \ch{H2+}, \ch{H2}, 
and \ch{H2-} as functions of interatomic distance. 

\subsection{Hydrogen molecule excited states}

We begin by analyzing the spin-excited states of the hydrogen molecule. In this 
context, spin-excited states contain also states with total spin $S=0$. Our previous 
works \cite{} demonstrated that diagonalizing the Hamiltonian in second quantization 
form results in six states: two singlets, one triplet and triple degenerated singlet.  
This outcome is achievable even with a basis set consisting solely of $1s$-type Slater
orbitals.

\begin{figure*}[t]
    \centering
    \includegraphics[width=1\textwidth]{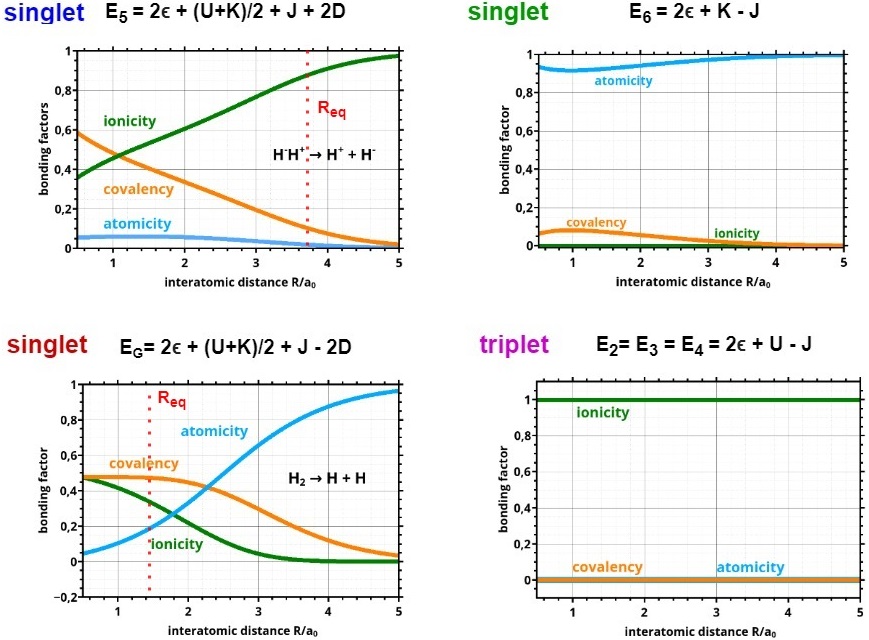}
    \caption{The atomicity, ionicity, and covalency for various ground and spin-excited states of the hydrogen molecule (\ch{H2}) as functions of the interatomic distance $R$. The energy expressions for these states include contributions from kinetic energy, Coulomb interaction, and exchange terms.
}
    \label{fig:levels_atom}
\end{figure*}

In particular, we investigate these excited states utilizing our concepts of true 
covalency, atomicity, and ionicity (bonding factors) \cite{}. To this end, we present 
in Fig.~\ref{fig:levels_atom} the bonding factors as functions of interatomic distance 
$R$ for all six states of the hydrogen molecule, with energy formulas indicated on 
these plots. The bonding characteristics for those states are shown in  Fig.~\ref{fig:levels_atom}.

The bottom-left panel shows the bonding 
characteristics for the ground state energy. As discussed earlier, the covalent 
contribution dominates at $R=R_{eq}$. However, there is a small admixture of 
atomic contribution. The ionic contribution is more noticeable than the atomic one, 
but as $R$ increases, both covalent and ionic characters vanish, while the atomic 
character persists. The proper eigenvector in the second quantization can be written 
in the following form

\begin{align}
    \ket{6} = C\left( \hat{a}^{\dagger}_{1\uparrow}\hat{a}^{\dagger}_{2\downarrow} + \hat{a}^{\dagger}_{1\downarrow}\hat{a}^{\dagger}_{2\uparrow} \right)\ket{0} \notag \\- I\left( \hat{a}^{\dagger}_{1\uparrow}\hat{a}^{\dagger}_{1\downarrow} + \hat{a}^{\dagger}_{2\uparrow}\hat{a}^{\dagger}_{2\downarrow} \right)\ket{0},
\end{align}

\noindent
where $C$ and $I$ as earlier denote covalent and ionic factors, respectively.  
The evolution from covalent-dominated bonding to the atomic limit is consistent with 
the intuitive picture resulting from the interpretation of the behavior of the 
binding energy curve, where in the limit $R \rightarrow \infty$ the molecule 
disperses into two individual atoms.

The top-left panel represents another one of the singlet states. In this case, 
ionicity contribution is dominant. The contribution of ionicity in this case is significantly smaller than for the ground state, which is obvious due to the nature of the wave function. As the interatomic distance increases covalency and atomicity as expected approach zero while ionicity reaches a value of one.

\begin{align}
    \ket{5} = C^{'}\left( \hat{a}^{\dagger}_{1\uparrow}\hat{a}^{\dagger}_{2\downarrow} + \hat{a}^{\dagger}_{1\downarrow}\hat{a}^{\dagger}_{2\uparrow} \right)\ket{0} \notag \\+I^{'}\left( \hat{a}^{\dagger}_{1\uparrow}\hat{a}^{\dagger}_{1\downarrow} + \hat{a}^{\dagger}_{2\uparrow}\hat{a}^{\dagger}_{2\downarrow} \right)\ket{0}
\end{align}

\noindent
in this case there are new functions $C^{'}$ and $I^{'}$ represent covalent and 
ionic factor, where $C^{'} = 4(t+V)/\sqrt{2D(D-U+K)}$ and $I^{'}=\sqrt{(D-U+K/2D)}$.
As shown in Fig.~\ref{fig:levels_atom}, this state has a small minimum at $R 
= 3.5a_0$. The molecule excited to this state has a metastable state with an ionic 
character and in the limit of infinite $R$, separated atoms occurs, but 
unlike for the ground state, it results in $\text{H}^+$ and $\text{H}^-$ ions 
rather than two neutral atoms.

The top-right panel also depicts the highest excited singlet state. In this state, 
atomicity is shown to be the dominant characteristic across the range of 
interatomic distances presented. Covalency and ionicity remain minimal, 
suggesting that the bonding in this excited state is predominantly characterized 
by atomicity. Three degenerated singlet eigenvectors are of the form 

\begin{align}
    \ket{1} &= \hat{a}^{\dagger}_{1\uparrow}\hat{a}^{\dagger}_{2\uparrow}\ket{0}, \\
    \ket{2} &= \hat{a}^{\dagger}_{1\uparrow}\hat{a}^{\dagger}_{2\downarrow}\ket{0}, \\
    \ket{3} &= \frac{1}{\sqrt{2}}\left( \hat{a}^{\dagger}_{1\uparrow}\hat{a}^{\dagger}_{2\downarrow} + \hat{a}^{\dagger}_{1\downarrow}\hat{a}^{\dagger}_{2\uparrow} \right)\ket{0}.
\end{align}

\noindent
In this case, covalency, ionicity and atomicity depend only on mixing parameters.

State $\ket{1}$ represents a configuration where both electrons are in a spin-up state. This kind of configuration typically does not contribute to the covalent bond because it does not allow for the formation of resonating paired electrons due to the Pauli exclusion, which is a prerequisite for strong covalent bonding.

The triplet $\ket{2}$ is characterized by the fact that one electron is in the spin-up state and one is in the spin-down state. It partially allows for covalent bonding, but given that the spins are not paired in the same orbital, the bonding is not as strong as in the fully paired configuration.

The bottom-right panel represents a triplet state. This panel shows that the 
ionicity is equal to one while covalency and atomicity remain zero. The eigenvector for this state is as follows

\begin{align}
    \ket{4} = \frac{1}{\sqrt{2}} \left( \hat{a}^{\dagger}_{1\uparrow}\hat{a}^{\dagger}_{1\downarrow} -
    \hat{a}^{\dagger}_{2\uparrow}\hat{a}^{\dagger}_{2\downarrow}\right)\ket{0}
\end{align}

\noindent
As in the previous case, bonding factors are only functions of mixing parameters.

One can see that in this wavefunction there is no covalent or atomic admixture. 
This state is also unstable, which leads to the conclusion that the molecule in 
this state is a kind of resonant structure between bound and free ionic forms.

Not only the wavefunction can be analyzed in the context of bonding characteristics, but energy can also be used for this purpose. Let us consider two similar cases of singlets: the ground state singlet and the excited singlet, the energy of which differs only by a factor of $2D$. In the first case, the energy contains a $+2D$ factor, and in the latter case, it is $-2D$.

The addition of $2D$ to the energy increases the contribution of the ionic 
interactions (related to $U$ and $K$). As $D$ is positive, adding it increases 
the separation of charges, leading to a more ionic character where the molecule 
dissociates into ions \ch{H^+} and \ch{H^-}. 

Covalent State (bottom left): The subtraction of $2D$ from the energy term 
reduces the contribution of the ionic interactions, enhancing the resonant nature of the bond where the electrons are more shared between the atoms. This leads to a lower 
energy state typical of covalent bonds, from the molecule dissociates into 
neutral atoms (\ch{H} + \ch{H}).

In the case of the singlet on the top-right side of the panel in 
Fig.~\ref{fig:levels_atom}, the atomic character based on the energy is quite 
obvious. The energy equation includes the term $K$, which generally represents 
interatomic Coulomb interaction promoting atomicity, and $-J$, which is an 
exchange interaction term that tends to destabilize the singlet state and is not 
present in the wavefunction. The presence of $K$ and $-J$ without any direct 
dependence on the intraatomic Coulomb interaction term $U$ indicates that this 
state is more likely to retain atomic characteristics rather than to form strong 
covalent or ionic bonds. Stronger atomicity in the graph aligns with this, as 
the energy expression supports a state where atoms remain relatively independent.

\subsection{Bonding as an energy competition}

It is a well-known fact, as demonstrated by Ruedenberg in his seminal work 
\cite{Ruedenberg1962}, that covalent bonding in hydrogen molecules occurs as a result 
of the kinetic energy decreasing with a simultaneous increase in the negative direction in potential energy. However, there are papers stating that this mechanism is not true in general and there are 
some molecular systems with different mechanisms lying on the bottom of the 
covalent bonding \cite{}. Regardless of the successes achieved in explaining the 
mechanisms of chemical bonding, the problem of ambiguity in results persists, as 
in the case of the \ch{LiH} molecule, where Molecular Orbital theory (MO) gives 
different results than Valence Bond theory (VB). This issue was the subject of 
research by M. Pendas et al. \cite{Pendas} and also part of our studies \cite{Hendzel2022}. 
Furthermore, methods based on energy decomposition are not without flaws, as they 
heavily depend on the numerical methods they are based on, which was discussed in 
this work \cite{EDA}. In this subsection, we propose an analysis of chemical bonding 
based on methods used in condensed matter physics, using the hydrogen molecule as 
an example.

To make this feasible, we start with the exact two-particle wave function, which was previously obtained by solving the Heitler-London-Slater model of the hydrogen molecule \cite{Hendzel2022}. Such wave function can be written as 
\begin{align}
\begin{split}
     &\Psi_0(\textbf{r}_1,\textbf{r}_2) = \frac{2(t+V)}{\sqrt{2D(D-U+K)}} \Psi_{cov}(\textbf{r}_1,\textbf{r}_2) \\
     &-\frac{1}{2}\sqrt{\frac{D-U+K}{2D}}\Psi_{ion}(\textbf{r}_1,\textbf{r}_2) \equiv \\ &\equiv  C\psi_{cov}(\textbf{r}_1,\textbf{r}_2) + I\psi_{ion}(\textbf{r}_1,\textbf{r}_2). 
\end{split}
     \label{coeff_wav1}
\end{align}

\noindent
In the given expression the wave function is expressed as a linear combination of 
two terms: $\psi_{cov}(\textbf{r}_1, \textbf{r}_2)$ and $\psi_{ion}(\textbf{r}_1, 
\textbf{r}_2)$. Covalent $C$ and ionic $I$ factors are functions of microscopic 
parameters introduced in the previous subsection. 

We consider the physical picture of why the chemical homopolar bond is covalent when the kinetic 
exchange interaction is higher than the effective Coulomb interaction. To do that we examine coefficients of the wave function from Eq.~\eqref{coeff_wav1}, where all parameters were explained but the role of $D$ can still be difficult to grasp. The expression $D = \sqrt{16(t+V)^2+(U-K)}$ determines the energy 
denominator of the wave function. Because our goal is to study the bonding factor as the ratio of kinetic energy to potential energy one can write it in the form
\begin{align}
&D = (U-K)\sqrt{1+\left(\frac{4(t+V)}{U-K}\right)^2} \notag\\
&=(U-K)\sqrt{1+\left(\frac{J_{kex}}{U-K}\right)} = \notag\\
&=E_{eff}\sqrt{1+\left(\frac{J_{kex}}{E_{eff}}\right)},
\end{align}

\noindent
where $D$ denominator is a function of kinetic exchange interaction $J_{kex}$ and 
effective Coulomb interaction $E_{eff}$. 

In the ionic regime, $(t+V)$ contribution should be smaller than $(U-K)$, whereas 
in the covalent regime, it is predominant. But, as one can expect, the ionic regime is unphysical as we consider the two-atomic homonuclear molecule. In the opposite case, $(t+V)$ contribution is of larger magnitude than $(U-K)$. Although, in general, the mechanism of covalent bond formation is consistent 
with that proposed by Reudenberg, there are fundamental differences between his and 
our solution. In Reudenberg's analysis, kinetic energy is the energy associated with 
the motion of electrons, whereas the exchange interaction energy is related to the virtually
resonating electrons in agreement with the understanding of covalent processes. The potential energy in Ruendenberg's analysis is calculated based on the 
electrostatic interactions between charged particles (nuclei and electrons) whereas in 
our case it is the difference between intraatomic and interatomic Coulomb repulsion, 
which is more natural to the discussion of ionic bonding. 

\begin{figure}[t]
    \centering
    \includegraphics[width=0.5\textwidth]{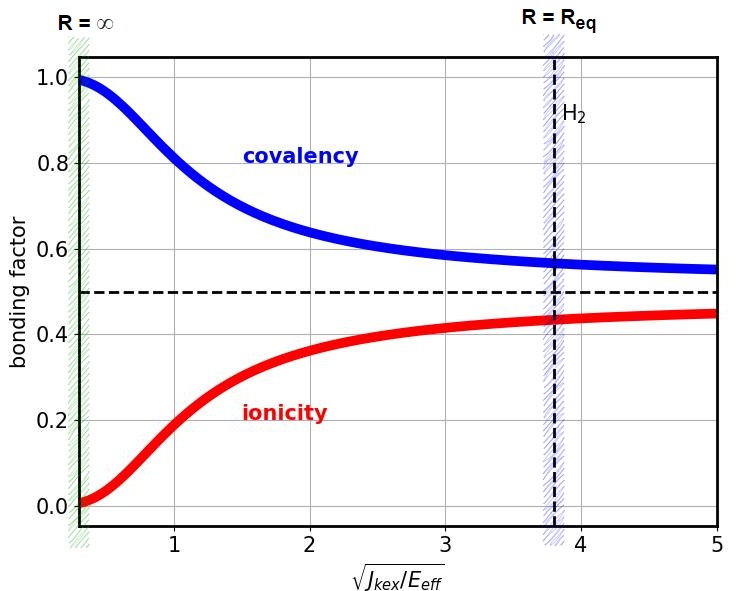}
    \caption{Evolution of the bonding factors: covalency and ionicity for \ch{H2} molecule with the square root of kinetic exchange and effective potential energy ratio. Two characteristic points are marked: the blue area is for that when $R=R_{eq}$, green area when $R\rightarrow \infty$.}
    \label{fig:j_vs_e}
\end{figure}

In Fig.~\ref{fig:j_vs_e} evolution of the two bonding factors, covalency and ionicity, is shown with the square 
root of the kinetic exchange to potential energy 
ratio. In our case, covalency dominates for all values of 
the ratio, especially when it is maximal and equal to unity. 
This feature appears naturally as this point corresponds to 
the separated atoms limit, and this is classical covalency (a 
mix of true covalency and atomicity, as we shown in 
\cite{hendzel2022toward}). What is new in our approach is that instead of decomposing the total energy into 
kinetic and potential components, we introduce effective parameters $J_{kex}$ and 
$E_{eff}$ based on the exact solution of the extended Heitler-London model for the 
hydrogen molecule.  It is also
possible to carry out it for more complex molecules, but the  
analysis may be cumbersome to implement.

\subsection{From one- to three-electron chemical bonding}

Starting from the hydrogen molecular cation \ch{H2+}, with only one electron, we can write second quantized wave functions as follows 

\begin{align}
    \ket{\psi_+^{\ch{H2+}}} = \frac{1}{\sqrt{2}}(\hat{a}^{\dagger}_1 + \hat{a}^{\dagger}_2 )\ket{0}, \\ 
    \ket{\psi_-^{\ch{H2+}}} = \frac{1}{\sqrt{2}}(\hat{a}^{\dagger}_1 - \hat{a}^{\dagger}_2)\ket{0}, 
\end{align}

\noindent
where only the bonding and antibonding orbitals appear. In this case, one can use 
relation from Eq.~\eqref{calc_wav} to calculate the corresponding Wannier functions of 
the form 

\begin{align}
    \psi_+^{\ch{H2+}}(\textbf{r}) = \frac{1}{\sqrt{2(1+S)}} (w_1(\textbf{r}) + w_2(\textbf{r})), \\
    \psi_-^{\ch{H2+}}(\textbf{r}) = \frac{1}{\sqrt{2(1+S)}} (w_1(\textbf{r}) - w_2(\textbf{r})), 
\end{align}

\noindent
which are transformed into the bonding molecular wave function to the atomic-based form
\begin{align}
    &\psi_-^{\ch{H2+}}(\textbf{r}) = \frac{\beta}{\sqrt{2(1+S)}} ((\phi_1(\textbf{r})-\phi_2(\textbf{r})) \notag \\&+\gamma (\phi_2(\textbf{r})-\phi_1(\textbf{r})).
\end{align}

In the case of the hydrogen molecule \ch{H2}, the ground state wave function can be 
written similarly as in Eq.~\eqref{coeff_wav1}.

In the case of \ch{H2-} ion, four states should be considered, leading to the 
construction of a $4\times4$ Hamiltonian, which yields two doubly degenerate states. 
For the lower energy state, the wave function in the language of second quantization 
takes the form

\begin{align}
    \ket{\psi^{\ch{H2-}}_G} = \frac{1}{\sqrt{2}} 
    (\hat{a}^{\dagger}_{1\uparrow}\hat{a}^{\dagger}_{2\uparrow}\hat{a}^{\dagger}_{2\downarrow
    } + \hat{a}^{\dagger}_{1\downarrow}\hat{a}^{\dagger}_{2\uparrow}\hat{a}^{\dagger}_{2\downarrow})\ket{0}.
\end{align}

\noindent
Equivalently this form has the following form in terms of the Wannier functions 

\begin{align}
    &\psi^{\ch{H2-}}_G(\textbf{r}_1, \textbf{r}_2, \textbf{r}_3) = \frac{1}{\sqrt{2}} (
    w_{1\uparrow}(\textbf{r}_1)w_{2\uparrow}(\textbf{r}_2)w_{2\downarrow}(\textbf{r}_3) \notag \\
    &+w_{1\downarrow}(\textbf{r}_1)w_{1\uparrow}(\textbf{r}_2)w_{1\uparrow}(\textbf{r}_3)). \label{three_wannier}
\end{align}

\noindent
It can be transformed to the three electron wave function consisting of linear 
combination of atomic functions
\begin{align}
    &\psi^{\ch{H2-}}_G(\textbf{r}_1, \textbf{r}_2, \textbf{r}_3) = \frac{\beta^3}{\sqrt{2}} (
    \phi_{1\uparrow}(\textbf{r}_1)\phi_{2\uparrow}(\textbf{r}_2)\phi_{2\downarrow}(\textbf{r}_3) \notag \\
    &+\phi_{1\downarrow}(\textbf{r}_1)\phi_{2\uparrow}(\textbf{r}_2)\phi_{2\uparrow}(\textbf{r}_3)  \notag \\
    &+\phi_{1\uparrow}(\textbf{r}_1)\phi_{1\downarrow}(\textbf{r}_2)\phi_{2\uparrow}(\textbf{r}_3)  \notag\\
    &+\phi_{1\uparrow}(\textbf{r}_1)\phi_{1\downarrow}(\textbf{r}_2)\phi_{2\downarrow}(\textbf{r}_3)
    ). \label{three_wannier}
\end{align}

We elaborate now on those three cases. The wave function for \ch{H2+} consists of  
partially covalent bonding where the single electron is spread over 
both nuclei, resulting in an intermediate bond strength. The bonding character is 
determined by both the overlap integral and mixing parameters ($\beta$, $\gamma$),
whereas in the case of \ch{H2}, it is a more complex function of microscopic ($t$, 
$U$, $V$, $K$), as well as the mixing parameters.

The wave function of \ch{H2} represents a stronger covalent bond case with an 
admixture of the ionic part, absent in the single-electron of \ch{H2+}.

The three-electron wave function (Eq.~\eqref{three_wannier}) reveals a more 
involved picture where distinguishing whether the chemical bond is covalent 
or ionic is not straightforward. It has a mixed resonant covalent-ionic bond character 
where two electrons always occupy one atom and the third is being transferred from one 
site to another. Such a process results in a more complex picture, in which the overall bond energy is reduced as shown in Tab.~\ref{table:comparison}.

\begin{table}[h]
\centering
\begin{tabular}{|c|c|c|c|}
\hline
Molecule & $R_{eq}$ ($a_0$) & \( E_{\text{bond}} \) (eV) & \( \frac{J_{kex}}{E_{eff}} \) \\
\hline
$\text{H}_2^{+}$ & 2.53 & -1.7721 & N/A \\
\hline
$\text{H}_2$ & 1.43 & -4.0749 & 3.86 \\
\hline
$\text{H}_2^-$ & 3.56 & -2.1792 & 0.22 \\
\hline
\end{tabular}
\caption{Comparison of bond lengths, $E_{\text{bond}}$, and $\frac{J}{E_{kex}}$ (on one particle) for \ch{H2+}, \ch{H2}, and \ch{H2-}.}
\label{table:comparison}
\end{table}

The presented values of bond energy and its length indicate that in either \ch{H2+} 
and \ch{H2-} ions are lower than in the case of \ch{H2} whereas the value of the ratio 
of effective kinetic energy to potential energy significantly larger than 1. 
Interestingly, the ratio for \ch{H2-} takes a value significantly lower than 1. We 
describe this reduction as the predominantly resonant character of the covalent-ionic 
bonding. 

\begin{figure}[t]
    \centering
    \includegraphics[width=0.5\textwidth]{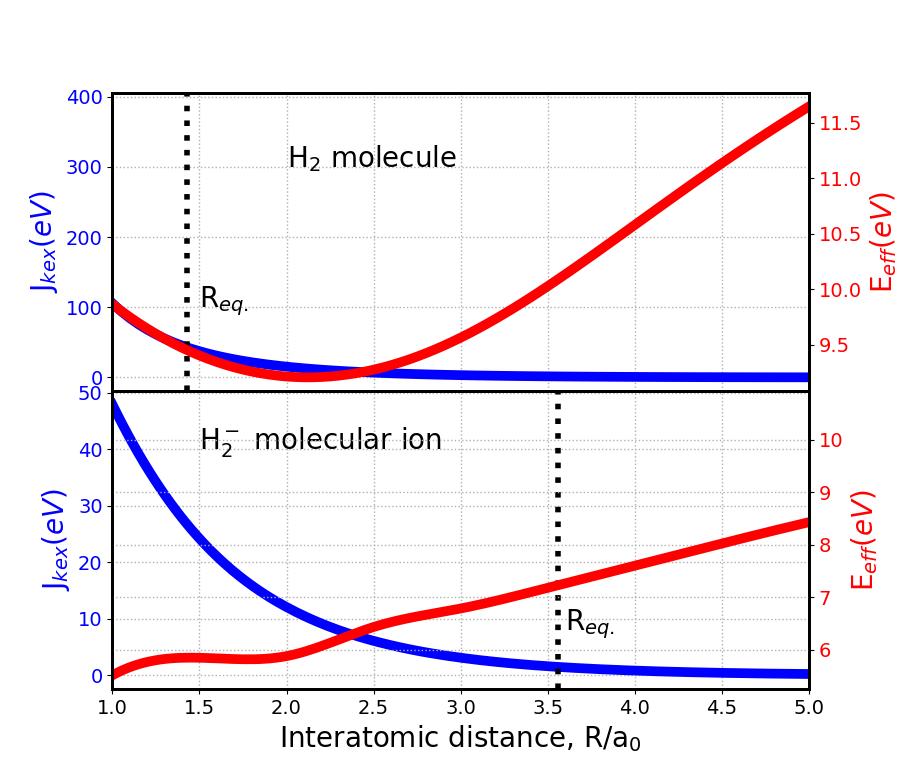}
    \caption{Comparison of kinetic-exchange energy (left y-axis) and effective Coulomb 
    interaction (right y-axis), both as a function of interatomic distance for \ch{H2} molecule (top) and \ch{H2^-} (bottom). Red curve in the bottom figure is not smooth, likely caused by numerical artifacts introduced during the calculations.}
    \label{fig:JE_mol}
\end{figure}

In the Fig.~\ref{fig:JE_mol} the kinetic-exchange energy and the effective 
Coulomb interaction, both as a function of interatomic distance $R$, is compared for 
\ch{H2} and \ch{H2-}. From the bonding perspective, the plots provide an insight into 
the nature of bonding in those molecules. The \ch{H2} molecule exhibits characteristics typical of covalent 
bonding, with the exchange interaction significantly higher than effective Coulomb 
interaction near the equilibrium distance (it is almost 4 times stronger). The effective kinetic energy decreases as the distance increases, still remaining larger than the effective potential energy. For the \ch{H2-} the 
exchange interaction energy decreases steadily with the increasing $R$
with the $J_{kex}/E_{eff}$ ratio less than 1 around $R_{eq}$. One should note that 
this relation does not relate to the ionic bonding character.

\begin{table}[h]
\centering
\caption{Comparison of key parameters: orbital size, ground state energy per one 
particle, and microscopic parameters at the points of minimal energy for \ch{H2} and 
\ch{H2-}, respectively.}
\begin{tabular}{|c|c|c|}
\hline
Parameter & \ch{H2} & \ch{H2-} \\
\hline
$E_G/N$ & -15.599 & -12.87 \\
$\alpha_0$ & 0.839 & 1.25/1.33 \\
$t$ & -9.905 & -1.487 \\
$U$ & 22.490 & 13.99 \\
$K$ & 13.007 & 6.705 \\
$V$ & -0.1578 & -0.1265 \\
$J$ & 0.2857 & 0.1238 \\
$\frac{4(t-V)^2}{(U-K)}$ & 37.39 & 1.604 \\
\hline
\end{tabular}
\label{values}
\end{table}

In Table~\ref{values} the basic quantities for \ch{H2} and \ch{H2-} are listed and 
compared at $R_{eq}$. For \ch{H2}, the ground state energy per particle is lower by about $3\text{ }eV$ as compared to \ch{H2-}. It means that the former is more stable. 

A smaller orbital size in the case of \ch{H2} suggests that the electrons on orbitals 
are more localized and closer to the nucleus. For the molecular ion, two $\alpha_0$ 
values are given since the sizes of the $1s$ orbitals are not the same as they are 
occupied by either one or two electrons. The hopping parameters for these two systems 
indicate that electron mobility is significantly higher for \ch{H2} (about four times 
higher). Additionally, Coulomb's repulsion magnitude is larger in the case of \ch{H2}.
Both interatomic and intraatomic repulsion values are approximately twice as 
high as the former. Similarly, the correlated hopping and exchange 
integrals are stronger in the neutral molecule.

Finally, the ratio $\frac{4(t-V)^2}{U-K}$ is 37.39 for \ch{H2} and 1.604 for 
\ch{H2-}, indicating a significantly stronger kinetic exchange interaction in the 
neutral molecule. This stronger interaction enhances effective electron sharing, 
promoting covalent bonding in \ch{H2}.

\section{Conlcusions}

We have analyzed here covalency, atomicity, and ionicity as the fundamental bonding characteristics for hydrogen-based systems: (\ch{H2+}, \ch{H2}, and \ch{H2-}).
The results provide a nuanced view of how bonds evolved between the different regimes: 
covalent, atomic, and ionic as the atoms move closer or separate. In this work, we tested and developed the conceptual framework that employs the formalism 
of combined first- and second-quantization. We discussed the exact solution of the 
extended Heitler-London model for the three systems which allowed us to precise 
quantification of bonding contributions coming from the kinetic exchange and effective 
Coulomb interactions.The analysis concentrates on the chemical bonding resulting 
from competition between those two energies. By examining these interactions 
\emph{vs.} $R$ we offer an intuitive understanding of bonding nature, and in particular, highlight the evolution from covalent to ionic character.

\section{acknowledgement}
This work was supported by Grants No.~UMO--2021/41/B/ST3/04070 and 2023/49/B/ST3/03545 
fromNarodowe Centrum Nauki.

\bibliography{hendzel}

\end{document}